\documentclass[aps,pre,preprint,superscriptaddress, showpacs]{revtex4}

\usepackage[dvips]{graphicx}
\begin{document}

\title{Parameter exploration of optically trapped liquid aerosols}

\author{D. R. Burnham}
\altaffiliation{Current address: Department of Chemistry, University of Washington, Box 351700, Seattle, WA 98195-1700}
\affiliation{SUPA, School of Physics and Astronomy, University of St Andrews, North Haugh, Fife, KY16 9SS, UK}
\affiliation{SUPA, Electronic Engineering and Physics Division, University of Dundee, Nethergate, Dundee, DD1 4HN, UK}

\author{P. J. Reece}
\altaffiliation{School of Physics, The University of New South Wales, Sydney, Australia 2052}
\affiliation{SUPA, School of Physics and Astronomy, University of St Andrews, North Haugh, Fife, KY16 9SS, UK}

\author{D. McGloin}
\affiliation{SUPA, Electronic Engineering and Physics Division, University of Dundee, Nethergate, Dundee, DD1 4HN, UK}

\date{\today}

\begin{abstract}
When studying the motion of optically trapped particles on the $\mu s$ time scale, in low viscous media such as air, inertia cannot be neglected. Resolution of unusual and interesting behaviour not seen in colloidal trapping experiments is possible. In attempt to explain the phenomena we use power spectral methods to perform a parameter study of the Brownian motion of optically trapped liquid aerosol droplets concentrated around the critically damped regime. We present evidence that the system is suitably described by a simple harmonic oscillator model which must include a description of Fax\'{e}n's correction, but not necessarily frequency dependent hydrodynamic corrections to Stokes' law. We also provide results describing how the system behaves under several variables and discuss the difficulty in decoupling the parameters responsible for the observed behaviour. We show that due to the relatively low dynamic viscosity and high trap stiffness it is easy to transfer between over- and under-damped motion by experimentally altering either trap stiffness or damping. Our results suggest stable aerosol trapping may be achieved in under-damped conditions, but the onset of deleterious optical forces at high trapping powers prevents the probing of the upper stability limits due to Brownian motion.
\end{abstract}


\maketitle

\section{Introduction}
In a number of recent publications our group~\citep{Burnham2009,DiLeonardo2007,McGloin2008} has presented an experimental system, based on the optical trapping of aerosols, for studying Brownian dynamics in both over- and under-damped conditions. With this unique perspective we are able to investigate the emergence of phenomena such as oscillatory motions due to the influence of inertial forces. This is a significant departure from traditional optical trapping experiments which are performed under conditions of heavy viscous damping~\citep{Berg-Sorensen2005}.

Its understanding~\citep{Einstein1956} can be used in conjunction with optical detection to determine the size of microscopic colloidal suspensions~\citep{Sun2001,Sudo2006} or to measure Avogadro's number~\citep{Newburgh2006}. Recently it has allowed optical traps to provide a powerful tool in diverse research fields capable of acting as a force transducer for molecular biology~\citep{Lang2002}, viscometry~\citep{Pesce2005}, microscopy~\citep{Rohrbach2004}, and fundamental physics~\citep{Hertlein2008}. These applications often use the power spectrum method~\citep{Allersma1998} to detect position~\citep{Denk1990}, measure forces~\citep{Ghislain1994}, or investigate colloidal dynamics~\citep{Meiners1999} and rely on the study of over-damped systems~\citep{Deng2007}.

The investigation of over-damped systems via optical trapping has produced classic experiments with important physical results including tests of Kramer's theory~\citep{McCann1999}, measurements of critical Casimir forces~\citep{Hertlein2008} and demonstrations of fluctuation theorems~\citep{Carberry2007}. The various optical potentials created through optical manipulation have, for example, been used to investigate colloidal crystals~\citep{Pertsinidis2001,Polin2006}, with particle dynamics providing analogies in thermal ratchets~\citep{Lee2005} and freezing~\citep{Chowdhury1985}.

All experiments in optical tweezers in a liquid environment behave as over-damped oscillators but there have been discussions that under-damped motions are observed~\citep{Joykutty2005} and comments that this should not be possible~\citep{Pedersen2007,Deng2007}. What is true is that studies of non over-damped systems are rare~\citep{DiLeonardo2007,Yao2009}. A recent resurgence in the original airborne particle experiments of Ashkin~\citep{Ashkin1971,Mitchem2008,McGloin2008,Guillon2009} looks set to change this, however, with opportunities to now study the little probed under-damped systems. The main applications of such experiments are in aerosol science with most studies to date investigating the chemistry of liquid droplets in gas phase environments~\cite{Butler2008,King2004} but they also offer more esoteric possibilities with opportunities to study quantum mechanical effects linked to Brownian motion \citep{Raizen2010}. Due to the importance of inertia in such systems they provide a drastically different damping environment and hence experimental possibilities.

When comparing aerosol trapping to aqueous trapping several phenomena may be observed that would be considered unusual. Varying trapping power alters the axial equilibrium position of droplets resulting in `power gradients'~\cite{Knox2007} with further increases causing their loss. Given a polydisperse nebulised sample the initial power used to capture a droplet has pronounced size selectivity~\cite{Hopkins2004,Burnham2006} and once trapped the droplet can undergo vertical oscillations at frequencies of $\sim0.1-10~\text{Hz}$.

This parameter study of the `mechanical' forces in airborne optical traps was conducted in an attempt to explain various phenomena but also proves useful in developing a deeper understanding of aerosol trapping.

A limited exploration has previously been carried out showing a trapped droplet can behave in either an over- or under-damped manner and that parametric resonance is easily excited~\cite{DiLeonardo2007}. In addition we have observed underdamped behavior in two particle systems~\cite{Yao2009}. This paper investigates the transition from over- to under-damped in more detail, exploring the parameter space by discussing dependence on laser power, droplet size, and depth into the sample at which the droplet is trapped. We show the system can be described by a simple harmonic oscillator model by including appropriate corrections. We also test the hypothesis that droplets are lost from their traps as the power is increased because they cross into the under-damped regime.

Several methods have been presented to characterise the Brownian dynamics of a trapped object, all relying on position measurement, and include, the drag force method, the equipartition method, the step response method, autocorrelation, and the power spectrum method~\cite{Neuman2004}. Position sensitive detectors, video tracking or quandrant photodiodes (QPDs) are often used to detect particle position each with their own speed and precision advantages and disadvantages~\cite{Crocker1996,Huisstede2005}. Here we employ a QPD due to their high bandwidth and use the power spectrum method, considered the most reliable~\cite{Volpe2007}, to characterise the Brownian motion of droplets within optical traps. The autocorrelation of a single particle is also feasible but can provide poor results in noisy systems. The power spectrum method decomposes the motion into frequency components so any noise can be easily dealt with. Although we are mainly concerned with observing the dynamics of trapped aerosols it is feasible to use this method to measure precise forces and position. Normally the method allows calculation of trap stiffness with prior knowledge of the viscosity of the surrounding medium and the particle radius~\cite{Neuman2004}, but it will be shown that when in air only the radius is needed.

First we will discuss the theory used to describe our experimental system along with any subtle corrections that may need to be considered and what the magnitude of their effect would be. We will then describe our experimental apparatus and procedures which differ slightly from optical trapping in aqueous media. We will present evidence that supports that our system is described by a simple harmonic oscillator model with certain corrections necessarily included. Finally we will try to determine the cause of the unusual behaviour partly by studying how trapped droplets behave near the critically damped regime.

\section{Theory}
Throughout this work we assume the particle velocity is well below the speed of sound and the propagations of interactions in the fluid are instantaneous, hence the fluid, air, is treated as incompressible~\cite{Landau1959}. An optically trapped particle is treated as residing in a harmonic potential well experiencing a Hookean restoring force when displaced through Brownian stochastic forces. The characteristic time for such a particle to lose energy through friction is a balance between inertial and viscous forces, $t_{inert} = m/\gamma_{0}$~\cite{Berg-Sorensen2004}, where $m$ is the particle mass and $\gamma_{0}$ is the viscous drag. For the smallest droplet studied here this time is longer than our experimental resolution so inertia must not be neglected as is usual for studies in liquid media. The Langevin equation describing the motion of a liquid aerosol of radius $R$, mass $m$, optically trapped in a fluid of temperature $T$, kinematic viscosity $\nu$, and density $\rho_{fluid}$, with stiffness $\kappa$ is~\cite{Wang1945}

\begin{equation}
{\ddot{x}(t)+\Gamma\dot{x}(t)+\Omega^{2}x(t) = \Lambda\zeta(t)},
\label{eq:1}
\end{equation}\newline
where $\Omega = \sqrt{(\kappa/m)}$ is the natural angular frequency of the droplet position fluctuations, $\Gamma = 6\pi\eta R/mC_{c}$ is the viscous damping of the medium due to a viscoscity $\eta$, $\Lambda = (2k_{B}T\Gamma/m)^{1/2}$~\cite{Chandrasekhar1943,Risken1989} is the Brownian stochastic force where $k_{B}$ is the Boltzmann constant, and for all $t$ and $t'$ $\langle\zeta(t)\rangle=0$ and $\langle\zeta(t)\zeta(t')\rangle=\delta(t-t')$. Stokes' Law is corrected for finite Knudsen number effects by including the empirical slip correction factor, $C_{c}$, with a $5.5\%-1.6\%$ reduction in drag for $3-10\mu\text{m}$ diameter droplets, respectively~\cite{Seinfeld1998}. Fourier transforming equation~\ref{eq:1} and finding the expectation value we decompose the motion into frequency components and find the power spectrum of position fluctuations to be

\begin{equation}
{S^{inertia}_{x}\left(\omega\right)=\frac{2k_{B}T}{\kappa}\frac{\Omega^{2}\Gamma}{\left(\omega^{2}-\Omega^{2}\right)^{2}+\omega^{2}\Gamma^{2}}},
\label{eq:2}
\end{equation}\newline
where $\omega$ is the angular frequency. This spectrum has a characteristic high frequency tail with $\omega^{-4}$ gradient and a plateau value at low frequencies equal to $2k_{B}T\Gamma/\kappa\Omega^{2}$. As inertia is included there is an additional limiting case, compared to over-damped oscillators, at the point of inflection equal to $2k_{B}T/\kappa\Gamma$. We define the ratio of damping coefficient to natural frequency as the `damping ratio', $\Gamma/\Omega$. For over-damped systems this is always greater than unity, as is found for colloidal systems where it is usually greater than ten. In such cases the first angular frequency term in the denominator can be neglected with respect to $\Omega$ to give the usual power spectrum for over-damped optical traps~\cite{DiLeonardo2007}. When trapping in air the system has the potential to become under-damped and hence $\Gamma/\Omega<1$.

The above Langevin equation assumes the motion occurs in bulk fluid media with uniform velocity, far away from other objects and surfaces so Stokes' law is only corrected for finite Knudsen number. However, in reality the objects here are undergoing linear harmonic motion within significant proximity ($\leq10R$) of a coverslip so it is inappropriate to assume Stokes' law still applies. Studying this problem it is seen there are two significant corrections that may need to be applied to the `in bulk' theory.

Firstly, the object is undergoing linear harmonic motion and so a more complex, frequency dependent, friction must be considered which was shown by Stokes to be~\cite{Berg-Sorensen2004,Stokes1850,Landau1959}

\begin{equation}
{\displaystyle F_{friction} = -\gamma_{0}\left(1+\sqrt{\frac{R^{2}\omega}{2\nu}}\right)\dot{x}}-\frac{2}{3}\pi\rho_{fluid}R^{3}\left(1+\frac{9}{2}\sqrt{\frac{2\nu}{R^{2}\omega}}\right)\ddot{x}.
\label{eq:3}
\end{equation}\newline
The first term comprises the familiar Stokes' drag plus a frequency dependent correction. The second term arises from the inertia created by any fluid entrained due to the past motion of the particle. This hydrodynamic correction is often neglected~\citep{Lang2002,DiLeonardo2007,Joykutty2005,Keen2007,Li2008}, at times with good cause, but needs to be applied when requiring precision $>10\%$~\citep{Berg-Sorensen2004}. Here we will try to justify our exclusion of such terms in a little detail.

Using equation~\ref{eq:3} and following Berg-S{\o}rensen and Flyvbjerg~\citep{Berg-Sorensen2004} we derive the hydrodynamically correct power spectrum in angular frequency to be
\begin{equation}
S_{x - hydro}^{inert} \left( \omega  \right) = \frac{{2k_B T}}{\kappa }\frac{{\Omega ^{2} \Gamma \left( {1 + \left( {\frac{\omega }{{\omega _\nu  }}} \right)^{1/2} } \right)}}{{\left( {\Omega ^{2}  - \Gamma \left( {\frac{{\omega ^{3/2} }}{{\omega _\nu ^{1/2} }}} \right) - \frac{{\omega ^{2} \Gamma }}{{\omega _m }}} \right)^{2}  + \left( {\omega \Gamma  + \Gamma \left( {\frac{{\omega ^{3/2} }}{{\omega _\nu ^{1/2} }}} \right)} \right)^{2} }},
\label{eq:4}
\end{equation}
where $\omega_{m} = \Gamma/(1+ \frac{2\pi \rho_{fluid} R^{3}}{3m}  ) $ and $\omega_{\nu} = 2\nu/R^{2}$. For systems incorporating inertia the usual definition of a corner frequency, $\omega_{c}=\kappa/\gamma_0$, clearly can no longer apply so the dependence has been removed. The low density of air reduces the denominator of $\omega_{m}$ to close to unity, effectively removing any effective mass considerations as the entrained fluid is negligible~\citep{Widom1971}. Decomposing equation~\ref{eq:3} into frequency components via Fourier theory (equation 31 in Berg-S{\o}rensen and Flyvbjerg~\citep{Berg-Sorensen2004}) shows the larger kinematic viscosity of air, hence $\omega_{\nu}$, reduces any correction to Stokes' law compared to trapping in water.

\begin{figure}[!ht]
\begin{centering}
\includegraphics[width=9cm]{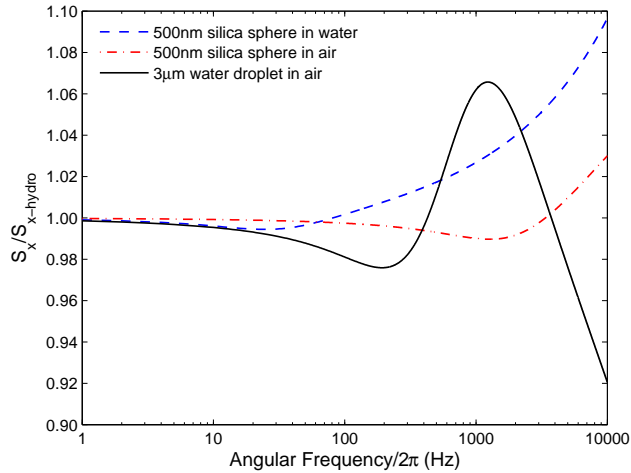}
\caption{Theoretical plot of $S_{x}/S_{x-hydro}$ as a function of angular frequency. For a given particle radius the hydrodynamic correction is smaller in air (red dot dashed) than in water (blue dashed), but, for the large liquid aerosols (black solid) the effect starts to become significant. Trap stiffness, $\kappa=2~\text{pN}\mu\text{m}^{-1}$.\label{fig:9}}
\end{centering}
\end{figure}

The effect on the hydrodynamic correction of a negligible effective mass and larger $\omega_{\nu}$ is not immediately apparent. To visualise the relative size of the correction we plot fig.~\ref{fig:9}; the ratio of the power spectrum in equation~\ref{eq:2} to the hydrodynamically correct version in equation~\ref{eq:4}. 

Clearly for a given particle type and size the correction is significantly smaller when studying aerosols. However, the solid line shows the error begins to become significant for aerosols with a radius that would be considered relatively large for particles normally used in power spectrum based studies in liquid. Should further studies be performed the hydrodynamic correction must be investigated to improve accuracy and precision.

For the majority of applications optical trapping parameters, such as trap stiffness, are only needed with an accuracy of $\sim10\%$, and considering the magnitude of the correction factor we believe it is reasonable to neglect the effect of frequency dependent friction in this study.

The second correction to be considered is that given by Fax\'{e}n regarding the force on a sphere in motion near a plane surface, exactly what occurs when trapping with high NA optical tweezers due to the proximity of coverslips. Here we only consider the correction in the lateral direction although both axial and rotational equivalents exist~\citep{Schaffer2007,Leach2009}. Fax\'{e}n's law shows the viscous drag on a sphere increases as it approaches a plane surface according to~\citep{Happel1965}
\begin{equation}
{\Gamma_{Faxen}=\frac{\Gamma}{1-\left(\frac{9R}{16L}\right)+\frac{1}{8}\left(\frac{R}{L}\right)^{3}-\frac{45}{256}\left(\frac{R}{L}\right)^{4}-\frac{1}{16}\left(\frac{R}{L}\right)^{5}}},
\label{eq:5}
\end{equation}
where $L$ is the distance between sphere centre and surface. For the particle sizes studied here this can have a dramatic effect on the friction experienced; even when trapping at distances approaching $40~\mu\text{m}$ from coverslips there can be a 7\% increase.

The final theoretical consideration is that to compare power spectra in given data sets we must calculate the detection system sensitivity, $\beta$, given in volts output per unit displacement of the particle. This is because the sensitivty can alter between experiments due to variations in power or simply geometry at the focus. Finding $\beta$ allows voltage power spectra, those recorded directly from the experiment, to be converted to physical spectra, $\text{nm}^{2}\text{Hz}^{-1}~\text{versus}~\text{Hz}$. Conventional methods rely on the relative simplicity of colloidal systems by using, for example, the drag force method~\citep{Malagnino2002}, its extension to an oscillating sample stage~\citep{Tolic2006} or moving a fixed bead over a known distance through the laser beam waist~\citep{Pralle1999}. Clearly the former two would be difficult to implement in air and the latter is obviously not a good replica of experimental conditions~\citep{Vermeulen2006}. A recent technique has been demonstrated that combines two methods to measure detector calibration from experimentally measured values alone~\citep{Tolic2006}. It is hoped, even with the unique problems of airborne trapping, by using AODs or SLMs to oscillate the trap position this technique will be developed for future experiments.

Here we are not concerned with high precision and for simplicity we calculate the detector sensitivity, $\beta$, from an uncalibrated voltage power spectrum $S^{V}(\omega) = \beta^{2}S_{inert}(\omega)$ using the plateau value, $P^{V}$, reached for $\omega\gg\Omega$ in the function $\omega^{4}S^{V}(\omega)$~\cite{Allersma1998}. We find the detector sensitivity, $\beta$, to be

\begin{equation}
\beta = \sqrt{\frac{P^{V}m}{2k_{B}T\Gamma}}.
\label{eq:6}
\end{equation}

\section{Experimental}
\begin{figure}
\begin{centering}
\includegraphics[width=8cm]{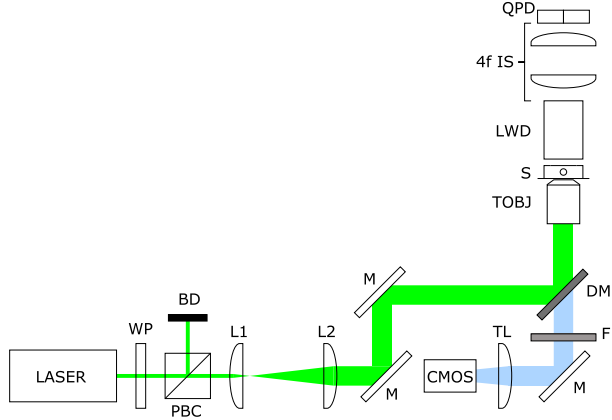}
\caption{Apparatus diagram. A Gaussian beam is expanded by lenses L1 and L2 and directed to slightly overfill the back aperture of the Nikon objective (TOBJ) with mirrors M and DM. The long working distance objective (LWD) collects the scattered light from the droplet and its back aperture is imaged onto the QPD via a 4f lens system. Power is controlled using a polarising beam cube (PBC) and half wave plate (WP). The same Nikon objective with an appropriate tube lens (TL) is used to image the sample (S) through a dichroic mirror (DM) and filter (F) onto the firewire camera (CMOS). The QPD, COBJ, and TOBJ are each mounted on three axis translation stages with the axial axis of TOBJ controlled either manually or by digital micrometer. BD is a beam dump.\label{fig:1}}
\end{centering}
\end{figure}

Droplets are trapped using a custom built inverted tweezers pictured in fig.~\ref{fig:1}. The beam from a 532nm Laser Quantum Finesse 4W c.w. laser is expanded by a Keplerian telescope to slightly overfill~\cite{Ashkin1992} the back aperture of a Nikon Plan 100x (NA = 1.25~\cite{fourtyfour}) oil immersion microscope objective. The beam is focused though a type one cover slip into an aerosol chamber constructed from a cylindrical plastic enclosure 9mm in height and 35mm in diameter. This produces an enclosed environment where a high relative humidity can exist and also shields the trapping region from external air currents. The top of the chamber is made from a type zero cover slip to allow for transmission and then collection of the scattered trapping laser by a long working distance (LWD) Mitutoyo 100x (NA = 0.55) objective, whose back aperture is imaged~\cite{Rohrbach2004} equally onto the four quadrants of a quadrant photodiode (QPD) (Hamamatsu Silicon Diode Array S5980) via a 4f imaging system. The Mitutoyo objective also acts as the condenser lens for K\"{o}hler illumination (not shown). The Nikon objective and an appropriate tube lens images the sample through a laser filter onto a Basler A602f firewire camera.

The liquid aerosol is produced by nebulising a salt solution (20-80g/L) with an Omron MicroAir NE-U22 vibrating mesh nebuliser which produces a polydisperse sample of liquid droplets with a mass median aerodynamic diameter of $4.9\mu\text{m}$~\cite{Omron}. The aerosol is transferred through a hole in the chamber side via a custom made tapered glass nozzle~\cite{Summers2008}.

\begin{figure}
\begin{centering}
\includegraphics[width=8cm]{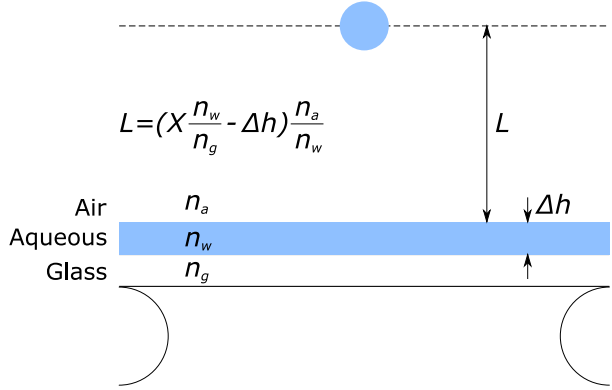}
\caption{Enlarged view of the trapping region in the sample of figure~\ref{fig:1}. The refractive indices of the coverslip and index matched oil, water, and air are $n_{g}$, $n_{w}$ and $n_{a}$ respectively. $\Delta h$ is the thickness of the water layer. Displacing the objective $X$ microns from being focussed on the first interface displaces the particle a distance $L$, given in the figure.\label{fig:2}}
\end{centering}
\end{figure}

The trapping beam is focused $\sim 30\mu\text{m}$ above the coverslips which are soaked in a 50\% aqueous dilution of `Decon 90' for longer than one week. This treatment increases the hydrophilicity of the glass and once aerosol has deposited on the cover slip it provides a relatively thin, flat, and uniform film of water above which we trap. De-ionised water saturated tissue paper is also placed in the chamber to increase the relative humidity, but we ensure it does not touch the cover slip as this can induce flows in the water layer. Figure~\ref{fig:2} shows an enlarged view of the trapping region geometry and also explains the relation between trapping height, \textit{L}, and objective displacement, \textit{X}.

Control over droplet size was required to fully investigate observed phenomena. Firstly, this was achieved imprecisely by varying the concentration of the nebulised salt solution~\cite{wet} as a higher concentration decreases the droplets vapour pressure allowing them to equilibrate with their surroundings at larger sizes. Secondly, more precise size selectivity can be induced with, on average, a positive linear dependence of captured droplet size on laser power~\cite{Hopkins2004,Burnham2006}.

Having trapped a droplet the nebuliser is turned off. Once the droplet has reached equilibrium with its surrounding environment, and the remaining aerosol settled, the current produced by the detection of light on the QPD is sent, via shielded cables, to amplification electronics~\citep{Pampaloni2002} containing a 50 kHz anti-aliasing filter. Data was acquired at a sampling frequency of 50 kHz for four seconds with a National Instruments PCI-6014E DAQ card, in differential mode. The voltage difference between left and right pairs of quadrants on the QPD represents the $x$ position and the difference between the top and bottom pairs represents the $y$ position. The voltage versus time data was Fourier transformed using LabVIEW and all remaining data analysis was performed offline at a later time. In order to minimise any parameter variation over time the experiments were carried out as quickly as possible with raw voltage versus time data not saved to increase speed still further. The detailed analysis of the data obtained for a colloidal case is extensively described in Berg-S{\o}rensen and Flyvbjerg~\citep{Berg-Sorensen2004}, and much remains the same here. An image of the trapped droplet was also taken with each power spectrum for later analysis.

To further reduce background noise work was always carried out, where possible, solitarily in the laboratory with the laser used at $>30\%$ capacity and power control achieved by using a pair of half wave plates with polarizing beam cubes. The first split the beam for two different experiments and the second controlled power for this experiment alone. The power was varied between a minimum of $0.702 \pm 0.009$mW and a maximum of $510 \pm 6$mW.

Unlike tweezing in water, simply increasing the trapping power does not assist in capturing an aerosol droplet from the nebulised cloud and as such initial laser power must be carefully selected. Each droplet trapped was subject to an increase in laser power in uniform steps with power spectra measurements taken at each. The minimum attainable damping ratio for each droplet was taken from the last power spectra measured before it fell from the trap upon increasing the power (i.e. the highest power). This represents an upper limit on the ratio for that size.

To study how the water-air interface to droplet height may affect the dynamics we simply keep a constant laser power and vary the height of the sample stage, controlled and measured by micrometer. The water layer thickness was measured by observing when a reflection of the trapping beam focus is obtained at both the water-air and glass-water interfaces. Having been focussed through two refractive index mismatched interfaces (glass to water and water to air) there will be an associated focal shift~\cite{Neuman2005} of which a rigorous description is complex~\cite{Torok1995,Torok1997} and not discussed here. A simple paraxial approximation is used to calculate the droplets position inside the chamber given a vertical displacement of the sample stage around a fixed objective. Modelling of the axial equilibrium position of the trap and experiments imaging the droplet from the side indicate that the relationship between droplet height and objective displacement is linear (data not shown) supporting the paraxial assumption~\cite{ModelFuture,Knox2007}.

\section{Results and Analysis}
\begin{figure}[!ht]
\begin{centering}
\includegraphics[width=9cm]{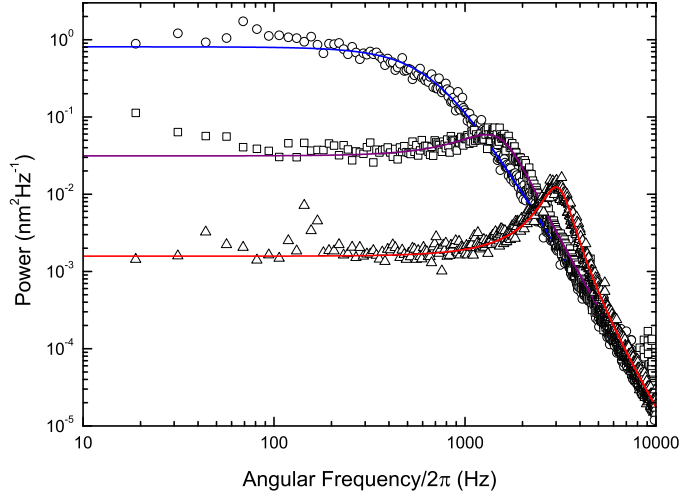}
\caption{Power spectra of a droplet of radius $3.7 \pm 0.2~\mu\text{m}$ trapped at powers $40.9 \pm 0.5$ (circles), $130 \pm 2$ (squares), and $356 \pm 4$ (triangles) mW resulting in damping ratios of $1.69 \pm 0.04$, $0.794 \pm 0.01$, and $0.364 \pm 0.001$ respectively. The fitting parameters for the top (blue), middle (purple) and bottom (red) curves are $\Omega/2\pi = 690 \pm 13, 1565 \pm 7, 3106 \pm 13~\text{Hz}$ and $\Gamma/2\pi = 1167 \pm 13, 1242 \pm 17, 1131 \pm 13~\text{Hz}$ respectively. As the power increases the appearance of a resonance peak is clear indicating the move into an under-damped regime, along with a decrease in area and hence position variance. The natural frequency increases with laser power because of the associated increase in lateral trap stiffness, $\kappa$.}
\label{fig:3}
\end{centering}
\end{figure}

Typical power spectra of position fluctuations from optically trapped droplets are shown in figure~\ref{fig:3}, illustrating, for a $3.7\pm0.2~\mu\text{m}$ radius droplet, the ease with which the system can be transferred between over- and under-damped dynamics by varying laser power. The tail falls off with $\omega^{-4}$ as expected for $\omega\gg\Omega$ from equation~\ref{eq:2} and a clear resonance peak begins to establish itself, indicative of the droplet moving through the critical and into the under-damped regime.

\begin{figure}[!ht]
\begin{centering}
\includegraphics[width=9cm]{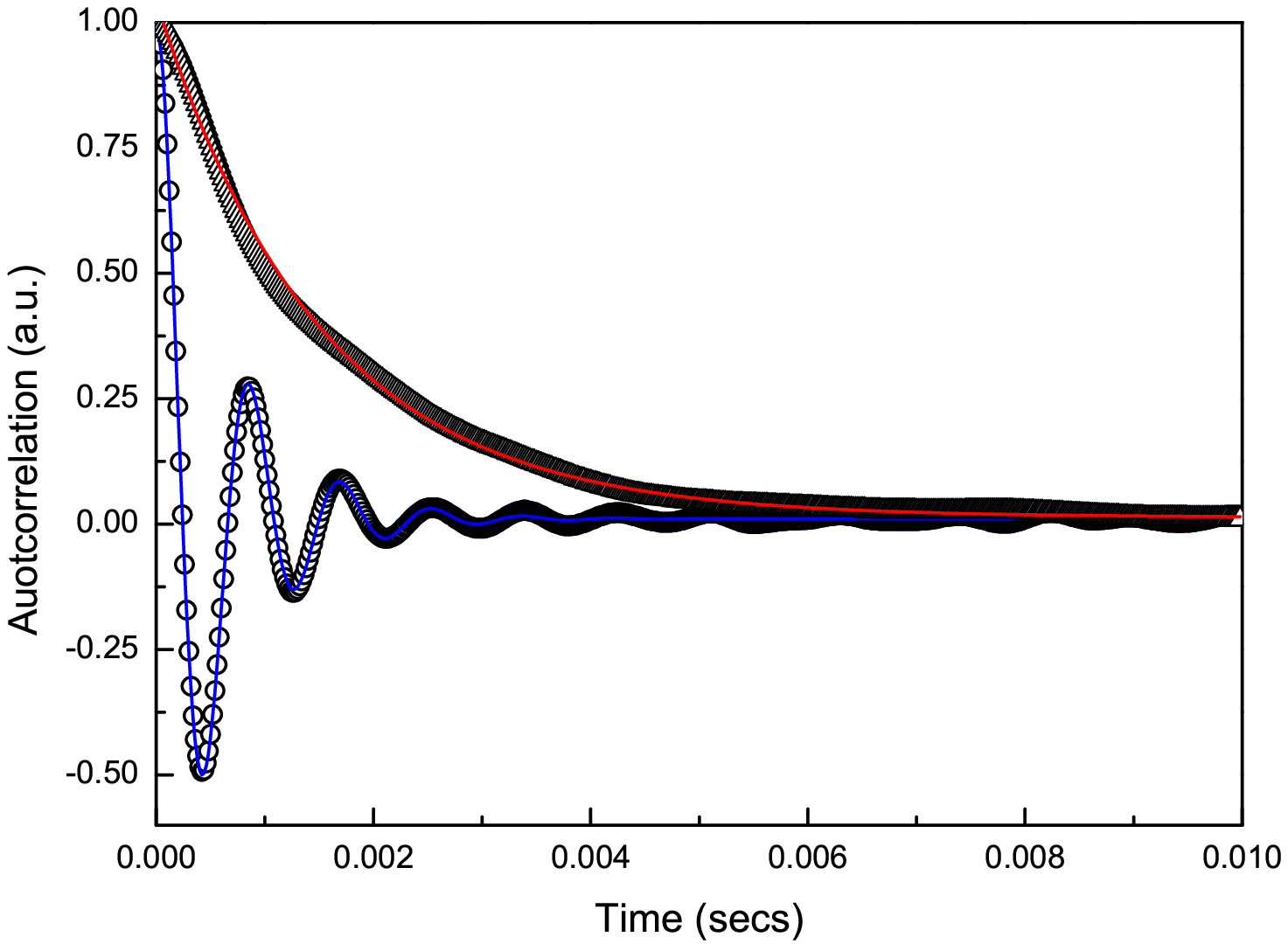}
\caption{Plot of experimental autocorrelation data with associated fits for a $5.2\pm0.2~\mu\text{m}$ optically trapped aerosol in an over- (triangles) and under-damped (circles) state trapped with powers $74\pm1$ and $442\pm7~\text{mW}$ respectively. Both traces clearly follow the classic exponential decay except in the under-damped case there is also the sinusoidal oscillation expected.}
\label{fig:AC}
\end{centering}
\end{figure}

For completeness, a plot of the autocorrelation function~\citep{Meiners1999} of a single droplet in an under- and over-damped state is shown in figure~\ref{fig:AC}. It shows the classic exponential decay for over-damped motion and sinusoidal oscillation enveloped by exponential decay for under-damped oscillators as expected.

The trend seen in figure~\ref{fig:3} remains for all droplets; an increase in power increases lateral trap stiffness and moves the system towards or into the under-damped regime. A range of damping ratios has been observed from $3.57 \pm 0.07$ down to $0.260 \pm 0.006$ over the $4.7 \pm 0.5~\mu\text{m}$ radius range studied. There is also an associated decrease in area under the power spectrum curve with increasing laser power, indicating a reduction in the position variance of the droplet.

The inclusion of inertial terms in the Brownian theory means only the mass of the particle is needed to calculate trap stiffness. Using the radius from video microscopy we obtain lateral trap stiffness values ranging from $0.12\pm0.10$ to $98\pm17~\text{pN}\mu\text{m}^{-1}$ for $1.0\pm0.3$ to $5.7\pm0.4~\mu\text{m}$ radius droplets.

\begin{figure}[!ht]
\begin{centering}
\includegraphics[width=9cm]{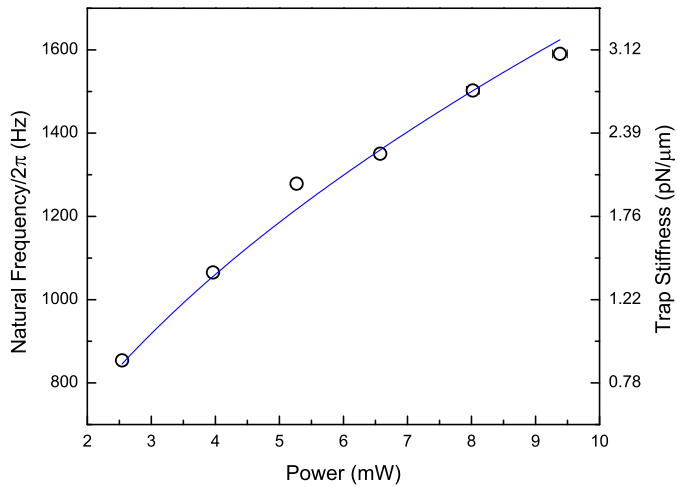}
\caption{An example of how natural frequency for a $1.8\pm0.2~\mu\text{m}$ radius droplet increases with the square root of laser power as expected from $\Omega = \sqrt{\kappa/m}$. The lateral trap stiffness axis is displayed for interest and is non linear. The error bars are standard error of the mean for the natural frequency rather than the trap stiffness (although they are smaller than the points themselves).\label{fig:4}}
\end{centering}
\end{figure}

One would expect the natural frequency of trapped droplets to vary as the square root of laser power, assuming the trap stiffness is linearly proportional to trapping power. This is confirmed in figure~\ref{fig:4} for a $1.8\pm0.2~\mu\text{m}$ radius droplet. For the range of radii and powers studied here we observe natural frequencies between $2\pi(328 \pm 12)~\text{rads}^{-1}$ and $2\pi(3433 \pm 15)~\text{rads}^{-1}$, falling close to and well above the corner frequencies measured by tweezers in liquid based systems, although obviously not directly comparable.

\begin{figure}[!ht]
\begin{centering}
\includegraphics[width=9cm]{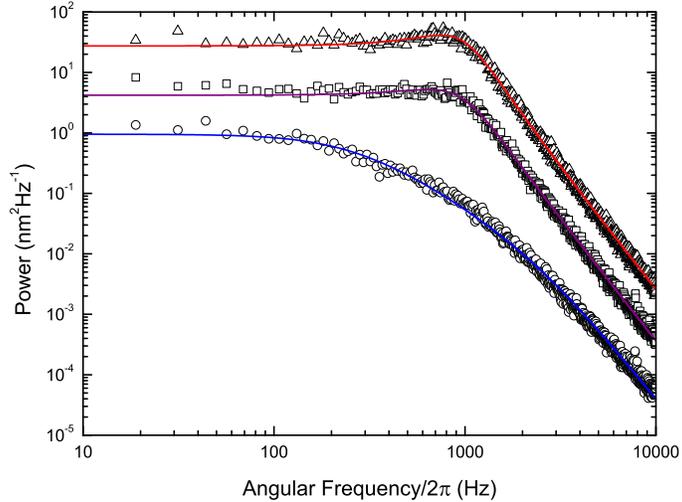}
\caption{Power spectra demonstrating changes in damping as a function of surface-droplet distance for a $3.8\pm0.2~\mu\text{m}$ radius droplet trapped with $46.3\pm0.6~\text{mW}$. The droplet was moved to heights of $4\pm1~\mu\text{m}$ (circles), $9\pm1~\mu\text{m}$ (squares), and $14\pm1~\mu\text{m}$ (triangles) above the water layer resulting in damping ratios of $3.40\pm0.06$, $1.06\pm0.01$, and $0.92\pm0.01$ respectively. The fitting parameters for the top, middle and bottom curves are $\Omega/2\pi = 981 \pm 5, 962 \pm 5, 815 \pm 10~\text{Hz}$ and $\Gamma/2\pi = 904 \pm 10, 1019 \pm 12, 2773 \pm 32~\text{Hz}$ respectively. The middle and top spectra are multiplied by 25 and 200 respectively to displace the data on the y-axis for clarity.\label{fig:6}}
\end{centering}
\end{figure}

The above results produce a downward shift in damping ratio by increasing lateral trap stiffness with larger laser powers. A decrease in friction felt by the droplet could likewise shift the ratio by varying the damping and as stated earlier Fax\'{e}n's correction predicts that the proximity of a surface to our microscopic object heavily influences this. Utilising this surface to droplet height dependence figure~\ref{fig:6} demonstrates that lowering the sample stage, hence increasing the distance, reduces the damping and transfers the system from over- to under-damped. Note the resonance peak remains approximately at the same frequency for each spectrum as only the damping is changing, not the trap stiffness, contrary to figure~\ref{fig:3}.

\begin{figure}
\begin{centering}
\includegraphics[width=8cm]{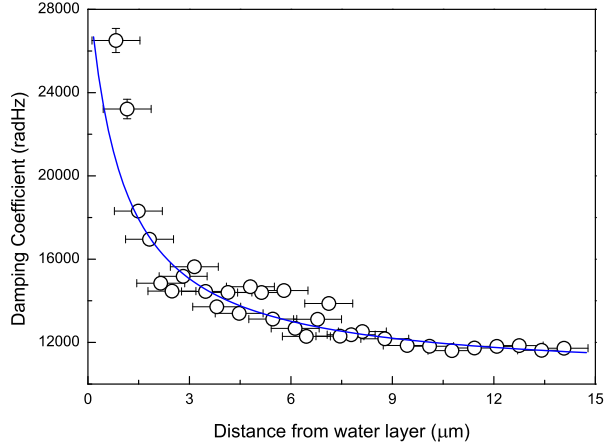}
\caption{Variation of damping experienced by a droplet, trapped with $6.2 \pm 0.1 mW$, as a function of surface-droplet height.\label{fig:7}}
\end{centering}
\end{figure}

Extracting damping values from data similar to fig.~\ref{fig:6} we can plot the dependence of friction upon droplet-surface height to obtain fig.~\ref{fig:7}~\cite{Burnham2009}. Here the micrometer raised the sample stage in increments of $1\mu\text{m}$, decreasing to $0.5\mu\text{m}$ as the surface was approached. 

\begin{figure}
\begin{centering}
\includegraphics[width=8cm]{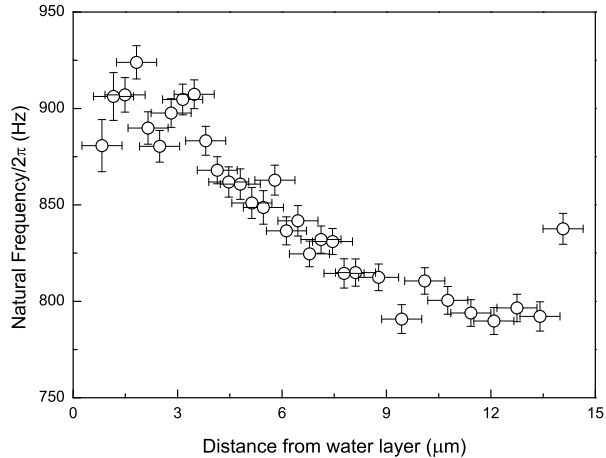}
\caption{Natural Frequency as a function of distance from water layer for the droplet in fig. 7.\label{fig:8}}
\end{centering}
\end{figure}

In fig.~\ref{fig:8} we plot the natural frequency as a function of height from the water layer. There is a steady fall off with distance indicating the spherical aberration induced is degrading the trap stiffness. Unlike the data of Vermeulen \textit{at al.}~\cite{Vermeulen2006} the data is approximately linear as the trap stiffness is independent of the viscous damping.

Figures~\ref{fig:3}-\ref{fig:4} and \ref{fig:8} confirm the simple harmonic oscillator model applies to our experimental system for all regimes of damping. Our results are not precise enough to examine the need for frequency dependent hydrodynamic correction, but figs.~\ref{fig:6} and~\ref{fig:7} do show Fax\'{e}n's correction has an important effect on the damping experienced.

\begin{figure}[!ht]
\begin{centering}
\includegraphics[width=9cm]{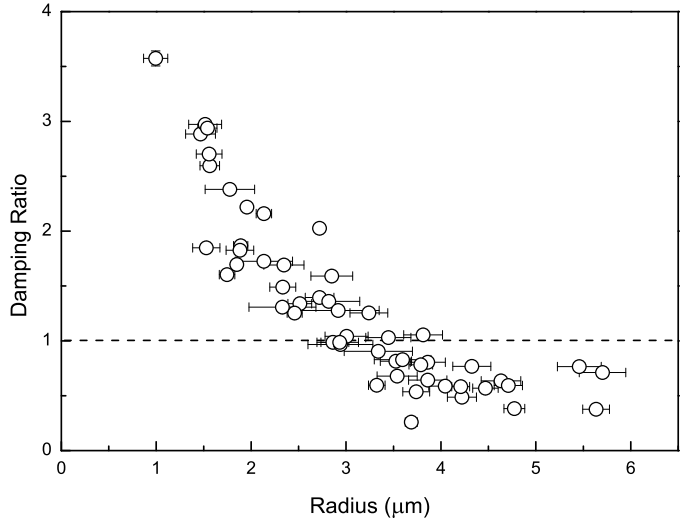}
\caption{Upper limit on the minimum attainable damping ratio against droplet radius. It is an upper limit as we increase the laser power in finite increments. The dashed horizontal line represents a critically damped system. The error bars are standard error of the mean.\label{fig:5}}
\end{centering}
\end{figure}

Bearing this in mind we turn to the most curious unusual phenomena observed; the loss of droplets from traps at a particular upper limit on laser power. The confirmation of a simple harmonic oscillator model as an appropriate description of our system leads us to test the hypothesis that the particles become largely under-damped so quickly that instability is caused. To test this and ascertain the true cause we plot, in figure~\ref{fig:5}, the upper limit on the minimum damping ratio attainable against droplet radius.

Figure~\ref{fig:5} illustrates droplets can exist in an under-damped regime. Some droplets are lost from traps while over-damped and some while under-damped. This suggests no instability is induced as the object crosses the critical damping point and so is not the reason for droplet loss with increasing trapping power. There is clear size dependence but one must be careful to note this does not lead to the conclusion that droplets do indeed become unstable as they reach a particular damping threshold, but rather fall from the optical traps at a certain upper limit on trapping power.


\section{Discussion}
The reader may notice significantly more scatter in the results of this investigation compared to experiments using similar techniques in liquid. The reasons will now be discussed briefly showing the complex nature of the experiment and the engineering challenges faced to improve future precision.

The majority of previous work using the power spectrum method is based on tweezing solid microspheres, with precisely known radii, in the liquid phase allowing very high precision studies; indeed, the ability to detect sphere non-uniformity is possible~\citep{Tolic2006}. In the studies here a large source of error is measuring the radius using video microscopy with the likely errors propagating heavily into some of the systems calculated properties ($\kappa\propto m\propto R^{3}$). Trapping of solid aerosols, with known radii would remove this problem but this is more difficult~\citep{Summers2008}. In addition, highly precise radius measurements are possible via CERS~\citep{Mitchem2006} but require sensitive spectrometers. We have shown in a previous publication that fitting to the data of fig.~\ref{fig:7} with equation~\ref{eq:5} allows the droplet radius to be measured with good precision~\cite{Burnham2009}.

It is difficult to determine which individual factor, trap stiffness or damping, contributes to the variation in damping ratio for any given experiment. For an individual droplet the surrounding conditions can remain relatively constant over the time of a single experiment as, with no additional aerosol flow from the nebuliser, the droplet quickly reaches equilibrium with its surroundings. To trap another droplet nebulisation must resume where upon the chamber conditions can alter. Additional aerosol can settle on the coverslip changing the thickness of the aqueous layer and hence the optical potential~\citep{Torok1997,ModelFuture,Viana2007} at the trap site together with the proximity of the particle to the surface~\citep{Berg-Sorensen2004,Happel1965,ModelFuture}. As mentioned variation in trapping power between droplets alters the height of the droplet~\citep{Knox2007} and hence distance from the underlying water layer, thus again altering the optical potential and damping. These factors contribute to the rather complex and difficult analysis of the system.

The difficult nature, relative to colloidal tweezers, of trapping in air imposes several important experimental methods. A long working distance condenser must be used due to the aerosol chamber height, but a higher NA lens may have been desirable to improve detector sensitivity~\citep{Rohrbach2003}. Most colloidal experiments use monodisperse suspensions of solid particles thus allowing an arbitrary number of measurement repetitions; often up to 100 power spectra are averaged. However, with the dynamic system investigated here (the droplets are continuously finding an equilibrium with the surrounding environment) the conditions of the experiment may not remain constant long enough for repeated measurements to improve precision, hence the choice of sampling and no averaging over multiple power spectra. Also, we are looking at an inherently unstable region with the aim, at times, of losing the trapped droplet so, clearly, another particle of the exact same size and composition cannot be found. With the current iteration of apparatus there is a clear trade off between speed and precision.

Some studies have used a secondary, independent probe beam to monitor position fluctuations as this allows greater flexibility and perhaps improved accuracy~\citep{Fallman2004}. We employ only a single beam because a very small amount of power is needed to tweeze in air~\citep{Burnham2006} and a second beam has the potential to significantly alter the potential at the trap site.

In future studies we suggest that a system including a `science chamber' be developed where many variables can be controlled. A particle could be trapped and transferred to such a chamber with relative humidity control, with or without a water layer, and with a lower physical profile to enable the use of higher NA condenser optics. Also the mechanical stability of our system is not fully optimised so the precision seen could be improved.

Not discussed in detail here is that Fax\'{e}n derived his correction for a sphere moving with constant velocity which is not the case, so for a complete solution the frequency dependent friction should be combined with Fax\'{e}n's correction~\citep{Berg-Sorensen2004,Tolic2006}. In addition when dealing with solid in fluid systems no slipping occurs at the boundary between the two materials upon translation. However, the physics involved becomes more complicated when studying fluid in fluid systems; slip can occur. Due to the possibility of slip at the surface of the fluid sphere, flow can be induced inside the water droplet. This flow causes reduction of the well known pre-factor of Stokes' law~\citep{Lamb1932,Happel1965} according to
\begin{equation}
	F_{stokes} = -\frac{6\pi\nu\rho Rv}{C_{c}}\frac{1+\frac{2}{3}\sigma}{1+\sigma},
\end{equation}
where $\sigma$ is the ratio of the dynamic viscosities of the medium and droplet, $\mu_{m}$ and $\mu_{p}$ respectively, giving Stokes' law for a water droplet in air, to be
\begin{equation}
	F_{stokes} = -\frac{5.96\pi\nu\rho Rv}{C_{c}},
\end{equation}
which has been taken into account in the data analysis here.

\section{Conclusion}
The work here is the first parameter exploration of the Brownian motion of optically trapped liquid aerosols. We have presented evidence that the system is suitably described by a simple harmonic oscillator model which must include a description of Fax\'{e}n's correction, but not necessarily frequency dependent hydrodynamic corrections to Stokes' law. The results also show there is difficulty in decoupling the parameters responsible for the observed behaviour. Having hyothesised that an instability is caused in the system when crossing from over- to under-damped regimes we see this is not supported by the evidence.

Considering the Langevin equation it is seen that there are only four processes providing forces that give rise to droplet position fluctuations; Brownian white noise, friction, inertia, and the optical force. Having rejected any damping or inertial cause for the instabilities with the evidence presented here the logical conclusion is that the optical force must determine whether the droplet remains trapped or not and gives rise to the size dependence of fig.~\ref{fig:5}. This will be discussed in a further publication.

The investigation has provided results extending the boundaries of precise studies of Brownian motion in optical tweezers into a new damping regime. It is hoped these results will provide researchers with a new understanding of optical tweezers for studies in both fundamental and applied science, providing a rich playground of study in the under-damped regime.


\begin{acknowledgments}
We thank EPSRC for funding this work. DRB would like to thank EPSRC for studentship support and is now a Lindemann Trust Fellow. DM is a Royal Society University Research Fellow.
\end{acknowledgments}


\begin{thebibliography}{0}
\expandafter\ifx\csname natexlab\endcsname\relax\def\natexlab#1{#1}\fi
\expandafter\ifx\csname bibnamefont\endcsname\relax
  \def\bibnamefont#1{#1}\fi
\expandafter\ifx\csname bibfnamefont\endcsname\relax
  \def\bibfnamefont#1{#1}\fi
\expandafter\ifx\csname citenamefont\endcsname\relax
  \def\citenamefont#1{#1}\fi
\expandafter\ifx\csname url\endcsname\relax
  \def\url#1{\texttt{#1}}\fi
\expandafter\ifx\csname urlprefix\endcsname\relax\def\urlprefix{URL }\fi
\providecommand{\bibinfo}[2]{#2}
\providecommand{\eprint}[2][]{\url{#2}}

\end{thebibliography}


\begin{thebibliography}{1}
\bibliographystyle{unsrt}

\bibitem{Burnham2009} D. R. Burnham and D. McGloin, New J. Phys., \textbf{11}, 063022 (2009).
\bibitem{DiLeonardo2007} R. Di Leonardo, G. Ruocco, J. Leach, M. J. Padgett, A. J. Wright, J. M. Girkin, D. R. Burnham, and D. McGloin, Phys. Rev. Lett. \textbf{99}, 010601 (2007).
\bibitem{McGloin2008} D. McGloin, D. R. Burnham, M. D. Summers, D. Rudd, N. Dewar and S. Anand, Faraday Discuss. \textbf{137}, 335 (2008).
\bibitem{Berg-Sorensen2005} K. Berg-Sorensen and H. Flyvbjerg, New J. Phys., \textbf{7}, 594-612 (2005).
\bibitem{Einstein1956} A. Einstein, ``Investigations on the Theory of the Brownian Movement'', Dover Publications (1956).
\bibitem{Sun2001} Z. G. Sun, C. D. Tomlin and E. M. Sevick-Muraca, Langmuir, \textbf{17} 6142-6147 (2001).
\bibitem{Sudo2006} S. Sudo, Y. Miyasaka, and K. Otsukam, Opt. Express, \textbf{14}, 1044-1054 (2006)
\bibitem{Newburgh2006} R. Newburgh, J. Peidleb and W. Ruecknerc, Am. J. Physics, \textbf{74}, 478-481 (2006)
\bibitem{Lang2002} M. J. Lang, C. L. Asbury, J. W. Shaevitz, and S. M. Block, Biophys. J. \textbf{83}, 491 (2002).
\bibitem{Pesce2005} G. Pesce, A. Sasso, and S. Fusco, Rev. Sci. Instrum. \textbf{76}, 115105 (2005).
\bibitem{Rohrbach2004} A. Rohrbach, C. Tischer, D. Neumayer, E. Florin, and E. H. K. Stelzer, Rev. Sci. Instrum. \textbf{75}, 2197 (2004).
\bibitem{Hertlein2008} C. Hertlein, L. Helden, A. Gambassi, S. Dietrich, and C. Bechinger, Nature, \textbf{451}, 172-175 (2008).
\bibitem{Allersma1998} M. W. Allersma, F. Gittes, M. J. deCastro, R. J. Stewart, and C. F. Schmidt, Biophys. J., \textbf{74}, 1074 (1998).
\bibitem{Denk1990} W. Denk and W. W. Webb, Appl. Opt. \textbf{29}, 2382 (1990).
\bibitem{Ghislain1994} L. Ghislain, N. Switz, and W. Webb, Rev. Sci. Instrum. \textbf{65}, 2762 (1994).
\bibitem{Meiners1999} J. Meiners and S. R. Quake, Phys. Rev. Lett., \textbf{82}, 2211 (1999).
\bibitem{Deng2007} Y. Deng, J. Bechhoefer, and N. R. Forde, J. Opt. A, \textbf{9}, S256--S263 (2007).
\bibitem{McCann1999} L. I. McCann, M. Dykman, and B. Golding, Nature, \textbf{402}, 785, (1999).
\bibitem{Carberry2007} D. M. Carberry, M. A. B. Baker, G. M. Wang, E. M. Sevick, and D. J. Evans, J. Opt. A, \textbf{9}, S204 (2007).
\bibitem{Pertsinidis2001} A. Pertsinidis and X. S. Ling, Phys. Rev. Lett., \textbf{87}, 098303 (2001).
\bibitem{Polin2006} M. Polin and D. G. Grier, Phys. Rev. Lett., \textbf{96}, 088101 (2006).
\bibitem{Lee2005} S. Lee and D. G. Grier, J. Phys.: Condens. Matter, \textbf{17}, S3685 (2005).
\bibitem{Chowdhury1985} A. Chowdhury, B. J. Ackerson and N. A. Clark, Phys. Rev. Lett., \textbf{55} 833-836 (1985)
\bibitem{Joykutty2005} J. Joykutty, V. Mathur, V. Venkataraman, and V. Natarajan, Phys. Rev. Lett. \textbf{95}, 193902 (2005).
\bibitem{Pedersen2007} L. Pedersen and H. Flyvbjerg, Phys. Rev. Lett. \textbf{98}, 189801 (2007).
\bibitem{Yao2009} A.M. Yao, S.A.J. Keen, D.R. Burnham, J. Leach, R. Di Leonardo, D. McGloin and M.J. Padgett, New J. Phys. \textbf{11} 053007 (2009)
\bibitem{Ashkin1971} A. Ashkin and J. M. Dziedzic, Appl. Phys. Lett. \textbf{19}, 283 (1971).
\bibitem{Mitchem2008} L. Mitchem and J. P. Reid, Chem. Soc. Rev. \textbf{37}, 756 (2008).
\bibitem{Guillon2009} M. Guillon, R. E. H. Miles, J. P. Reid and D. McGloin, New J. Phys. \textbf{11}, 103041 (2009).
\bibitem{Butler2008} J. R. Butler, L. Mitchem, K. L. Hanford, L. Treuel, and J. P. Reid, Faraday Discuss, 137, 351-366 (2008).
\bibitem{King2004} M. D. King, K. C. Thompson, and A. D. Ward J. Am. Chem. Soc., \textbf{126}, 16710-16711 (2004).
\bibitem{Raizen2010} T. Li, S. Kheifets, D. Medellin and M. G. Raizen, Sciencexpress (2010) doi:10.1126/science.1189402
\bibitem{Knox2007} K. J. Knox, J. P. Reid, K. L. Hanford, A. J. Hudson, and L. Mitchem, J. Opt. A, \textbf{9}, S180 (2007).
\bibitem{Hopkins2004} R. J. Hopkins, L. Mitchem, A. D. Ward, and J. P. Reid, Phys. Chem. Chem. Phys., \textbf{6}, 4924-4927 (2004).
\bibitem{Burnham2006} D. R. Burnham and D. McGloin, Opt. Express, \textbf{14}, 4175 (2006).
\bibitem{Neuman2004} Keir C. Neuman and Steven M. Block, Rev. Sci. Instrum. \textbf{75}, 2787-2809 (2004).
\bibitem{Crocker1996} J. C. Crocker and D. G. Grier, J. Colloid and Interface Science, \textbf{179}, 298-310 (1996).
\bibitem{Huisstede2005} J. H. G. Huisstede, K. O. van der Werf, M. L. Bennink, V. Subramaniam, Optics Express, \textbf{13}, 1113-1123 (2005).
\bibitem{Volpe2007} G. Volpe, and D. Petrov, Phys. Rev. E, \textbf{76}, 061118 (2007).
\bibitem{Landau1959} ``Fluid Mechanics'', L. D. Landau and E. M. Lifshitz, Pergamon Press (1959).
\bibitem{Berg-Sorensen2004} K. Berg-S{\o}rensen and H. Flyvbjerg, Rev. Sci. Instrum. \textbf{75}, 594 (2004).
\bibitem{Wang1945} M. C. Wang and G. E. Uhlenbeck, Rev. Mod. Phys. \textbf{17}, 323 (1945).
\bibitem{Chandrasekhar1943} S. Chandrasekhar, Rev. Mod. Phys. \textbf{15}, 1 (1943).
\bibitem{Risken1989} H. Risken, ``The Fokker-Plank Equation: Methods of Solution and Applications'', 2nd Edition Springer (1989).
\bibitem{Seinfeld1998} J. H. Seinfeld and S. N. Pandis, ``Atmospheric Chemistry and Physics: From Air Pollution to Climate Change'', Wiley-Interscience (1998).
\bibitem{Stokes1850} G. G. Stokes, ``On the effect of the internal friction of fluids on the motion of pendulums'', Cambridge Philosophical Society Transactions \textbf{IX} (1850).
\bibitem{Keen2007} S. Keen, J. Leach, G. Gibson, and M. J. Padgett, \textbf{9}, S264 (2007).
\bibitem{Li2008} M. Li and J. Arlt, \textbf{281}, 135-140 (2008).
\bibitem{Widom1971} A. Widom, Phys. Rev. A, \textbf{3}, 1394-1396 (1971).
\bibitem{Schaffer2007} E. Schaffer, S. F. Norrelykke and J. Howard, Langmuir, \textbf{23} 3654-3665 (2007)
\bibitem{Leach2009} J. Leach, H. Mushfique, S. Keen, R. Di Leonardo, G. Ruocco, J. M. Cooper and M. J. Padgett, Phys. Rev. E, \textbf{79} 026301 (2009)
\bibitem{Happel1965} J. Happel and H. Brenner, Prentice Hall (1965).
\bibitem{Malagnino2002} N. Malagnino, G. Pesce, A. Sasso, and E. Arimondo, Optics Communications, \textbf{214}, 15--24 (2002).
\bibitem{Tolic2006} S. F. Tolic-N{\o}rrelykke, E. Schäffer, J. Howard, F. S. Pavone, F. Juelicher, H. Flyvbjerg, Rev. Sci. Instrum. \textbf{77}, 103106 (2006).
\bibitem{Pralle1999} A. Pralle, M. Prummer, E. L. Florin, E. H. K. Stelzer and J. K. H. H\"{o}rber, Microscopy Research and Techniques, \textbf{44}, 378-386 (1999).
\bibitem{Vermeulen2006} K. Vermeulen, G. Wuite, G. Stienen and C. Schmidt, Appl. Opt. \textbf{45}, 1812-1819 (2006).
\bibitem{Ashkin1992} A. Ashkin, Biophysical J. \textbf{61}, 569 (1992).
\bibitem{fourtyfour} It is important to note that the NA can clearly not be larger than unity in the focal region and in fact due to total internal reflection at the glass:water:air boundary the NA is effectively reduced to $\sim0.67$.
\bibitem{Omron} Datasheet accompanying Omron MicroAir NE-U22 nebuliser
\bibitem{Summers2008} M. D. Summers, D. R. Burnham, and D. McGloin, Opt. Express, \textbf{16}, 7739-7747 (2007).
\bibitem{wet} J. H. Seinfeld and S. N. Pandis, "Atmospheric chemistry and physics: Air pollution to climate change"
John Wiley and Sons Inc. (1998).
\bibitem{Pampaloni2002} F. Pampaloni ``Force sensing and surface analysis with optically trapped microprobes'' Universit\"{a}t Regensburg (2002)
\bibitem{Neuman2005} K. C. Neuman, E. A. Abbondanzieri, and S. M. Block, Opt. Lett., \textbf{30}, 1318-1320 (2005).
\bibitem{Torok1995} P. T\"{o}r\"{o}k, P. Vagra, Z. Laczik, and G. R. Brooker, J. Opt. Soc. Am. A, \textbf{12}, 325-332 (1995).
\bibitem{Torok1997} P. T\"{o}r\"{o}k, P. Vagra, Appl. Opt., \textbf{36}, 2305-2312 (1997).
\bibitem{ModelFuture} D. R. Burnham and D. McGloin, ``Modelling of optical traps for aerosols'', submitted.

\bibitem{Mitchem2006} L. Mitchem, J. Buajarern, R. J. Hopkins, A. D. Ward, R. J. J. Gilham, R. L. Johnston, and J. P. Reid, J Phys. Chem. A, \textbf{10}, 8116 (2006).
\bibitem{Viana2007} N. B. Viana, M. S. Rocha, O. N. Mesquita, A. Mazolli, P. A. Maia Neto, and H. M. Nussenzveig, Phys. Rev. E, \textbf{75}, 021914 (2007).
\bibitem{nm} K J Knox and J P Reid, J. Phys. Chem. A, \textbf{112} 10439–-10441 (2008).
\bibitem{Rohrbach2003} A. Rohrbach, H. Kress and E. H. K.Stelzer, Opt. Lett., \textbf{28}, 411-413 (2003)
\bibitem{Fallman2004} E. F\"{a}llman, S. Schedin, J. Jass, M. Andersson, B. E. Uhlin and O. Axner, Biosensors and Bioelectronics, \textbf{19}, 1429-1437 (2004)
\bibitem{Lamb1932} H. Lamb, ``Hydrodynamics'', Cambridge (1932).




\end{thebibliography}
\end{document}